\documentclass[
    ,draft            
  ]
  {aipproc}

\layoutstyle{6x9}

\newcommand{\Mdot}{\dot{M}}

\newcommand{\Msun}{M_{\odot}}
\newcommand{\Rsun}{R_{\odot}}
\newcommand{\yr}{\rm yr^{-1}}
\newcommand{\ion}[2]{#1$\,${\small{#2}}\relax}
\newcommand{\lax}{{\lower0.75ex\hbox{ $<$ }\atop\raise0.5ex\hbox{ $\sim$ }}}
\newcommand{\gax}{{\lower0.75ex\hbox{ $>$ }\atop\raise0.5ex\hbox{ $\sim$ }}}

\begin{document}

\title{The Physics of Wind-Fed Accretion}



\author{Christopher W.~Mauche}{
  address={Lawrence Livermore National Laboratory,
   L-473, 7000 East Avenue, Livermore, CA 94550},
}

\author{Duane A.~Liedahl}{
  address={Lawrence Livermore National Laboratory,
   L-473, 7000 East Avenue, Livermore, CA 94550},
}

\author{Shizuka Akiyama}{
  address={KIPAC, Stanford University,
    2575 Sand Hill Road, M/S 29, Menlo Park, CA 94025},
}

\author{Tomasz Plewa}{
  address={Florida State University,
    Department of Scientific Computing, DSL 443, Tallahassee, FL 32306}
}

\begin{abstract}

We provide a brief review of the physical processes behind the radiative driving 
of the winds of OB stars and the Bondi-Hoyle-Lyttleton capture and accretion of 
a fraction of the stellar wind by a compact object, typically a neutron star, in
detached high-mass X-ray binaries (HMXBs). In addition, we describe a program to 
develop global models of the radiatively-driven photoionized winds and accretion 
flows of HMXBs, with particular attention to the prototypical system Vela X-1. 
The models combine XSTAR photoionization calculations, HULLAC emission models
appropriate to X-ray photoionized plasmas, improved models of the radiative 
driving of photoionized winds, FLASH time-dependent adaptive-mesh hydrodynamics 
calculations, and Monte Carlo radiation transport. We present two- and
three-dimensional maps of the density, temperature, velocity, ionization parameter, 
and emissivity distributions of representative X-ray emission lines, as well as 
synthetic global Monte Carlo X-ray spectra. Such models help to better constrain
the properties of the winds of HMXBs, which bear on such fundamental questions 
as the long-term evolution of these binaries and the chemical enrichment of the 
interstellar medium.

\end{abstract}

\maketitle


\section{Introduction}

As described by Castor, Abbott, and Klein (hereafter CAK)
\cite{cak75}, mass loss in the form of a high-velocity wind is driven
from the surface of OB stars by radiation pressure on a multitude of
resonance transitions of intermediate charge states of cosmically
abundant elements.
The wind is characterized by a mass-loss rate
$\Mdot\approx 10^{-6}$--$10^{-5}~\Msun~\yr $ and a velocity profile
$v(r)\approx v_\infty (1-R_\star /r)^\beta $, where
$r$ is the distance from the OB star,
$\beta \approx \frac{1}{2}$,
$v_\infty\approx 3 v_{\rm esc}$ is the terminal velocity,
$v_{\rm esc}=(2GM_\star /R_\star )^{1/2}$ is the escape velocity, and
$M_\star $ and $R_\star $ are respectively the mass and radius of the
OB star. As described by Hoyle and Lyttleton \cite{hoy39}, Bondi
\cite{bon52}, and Shapiro and Lightman \cite{sha76}, in detached
high-mass X-ray binaries (HMXBs), a compact object, typically a
neutron star, captures the fraction of the wind
of the OB star that passes within the accretion radius
$r_{\rm a} = \zeta 2GM_{\rm x}/v_{\rm rel}^2$ of the compact object, where
$\zeta $ is a constant of order unity,
$v_{\rm rel} = (v_{\rm w}^2+v_{\rm x}^2)^{1/2}$
is the velocity of the wind relative to the compact object,
$v_{\rm w}$ is the velocity of the wind at $r=a$,
$v_{\rm x} = 2\pi a/P_{\rm orb}$ is the orbital velocity of the compact object,
$a=[G(M_\star +M_{\rm x} )(P_{\rm orb}/2\pi )^2]^{1/3}$ is the binary
separation, and $P_{\rm orb}$ is the binary orbital period. The
mass-accretion rate onto the compact object
$\Mdot_{\rm a} = \pi r_{\rm a}^2\rho _{\rm w}v_{\rm rel}$, where
$\rho _{\rm w} = \Mdot_{\rm w}/4\pi a^2v_{\rm w}$ is the wind mass density and
$\Mdot_{\rm w}$ is the wind mass-loss rate, hence
$\Mdot_{\rm a} =
\frac{1}{4}\,
(r_{\rm a}/a)^2
(v_{\rm rel}/v_{\rm w}) \Mdot_{\rm w}\equiv f \Mdot_{\rm w}$.
This compact expression expands out to
\begin{equation}
\Mdot_{\rm a}\approx
0.012\, \zeta^2 \eta^{-4}
(M_{\rm x}/\Msun )^2
(M_\star/\Msun )^{-8/3}
(R_\star/\Rsun )^2
(P_{\rm orb}/{\rm d})^{-4/3} \Mdot_{\rm w},
\end{equation}
where
$\eta\equiv (v_{\rm w}/v_{\rm esc})\approx
3\, (1-R_\star/a)^\beta$ for an isolated OB star.
For the parameters appropriate to the prototypical detached HMXB Vela X-1
(\S 2),\footnote{$M_\star =23.8~\Msun $,
$R_\star =30~\Rsun $,
$M_{\rm x}=1.86~\Msun $,
$P_{\rm orb}=8.964$ d,
$v_\infty = 1700~\rm km~s^{-1}$,
$\beta=0.8$, and
$\Mdot _{\rm w}\approx 10^{-6}~\Msun~\yr $, hence
$a=3.7\times 10^{12}$ cm,
$r_{\rm a} = 5.7\times 10^{10}$ cm,
$v_{\rm x}=300~\rm km~s^{-1}$,
$v_{\rm w}=880~\rm km~s^{-1}$,
$v_{\rm rel}=930~\rm km~s^{-1}$,
$v_{\rm esc}=550~\rm km~s^{-1}$, and
$\eta = 1.6$.}
$\Mdot_{\rm a}\approx 7\times 10^{-5} \Mdot_{\rm w}\sim
7\times 10^{-11}~\Msun~\yr $.
Accretion of this material onto the neutron star powers an X-ray luminosity
$L_{\rm x} = G\Mdot_{\rm a}M_{\rm x}/R_{\rm x}\sim
1\times 10^{36}~{\rm erg~s}^{-1}$, where
$M_{\rm x}$ and $R_{\rm x}$ are respectively the mass and radius of
the compact object. The resulting X-ray flux photoionizes the wind and
reduces its ability to be radiatively driven, both because the higher
ionization state of the plasma results in a reduction in the number of
resonance transitions, and because the energy of the transitions
shifts to shorter wavelengths where the overlap with the stellar
continuum is lower. To first order, the lower radiative driving
results in a reduced wind velocity $v_{\rm w}$ near the compact
object, which increases the accretion radius $r_{\rm a}$, which
increases the accretion efficiency $f$, which increases the X-ray
luminosity $L_{\rm x}$. In this way, the X-ray emission of HMXBs is
the result of a complex interplay between the radiative driving of the
wind of the OB star and the photoionization of the wind by the neutron
star.

Known since the early days of X-ray astronomy, HMXBs have been extensively
studied observationally, theoretically \cite{hat77,mcc84,ste90}, and
computationally \cite{blo90,blo91,blo94,blo95}. They are excellent
targets for X-ray spectroscopic observations because the large covering
fraction of the wind and the moderate X-ray luminosities result in large
volumes of photoionized plasma that produce strong recombination lines and
narrow radiative recombination continua of H- and He-like ions, as well as
fluorescent lines from lower charge states.

\begin{figure}
\includegraphics[scale=0.875]{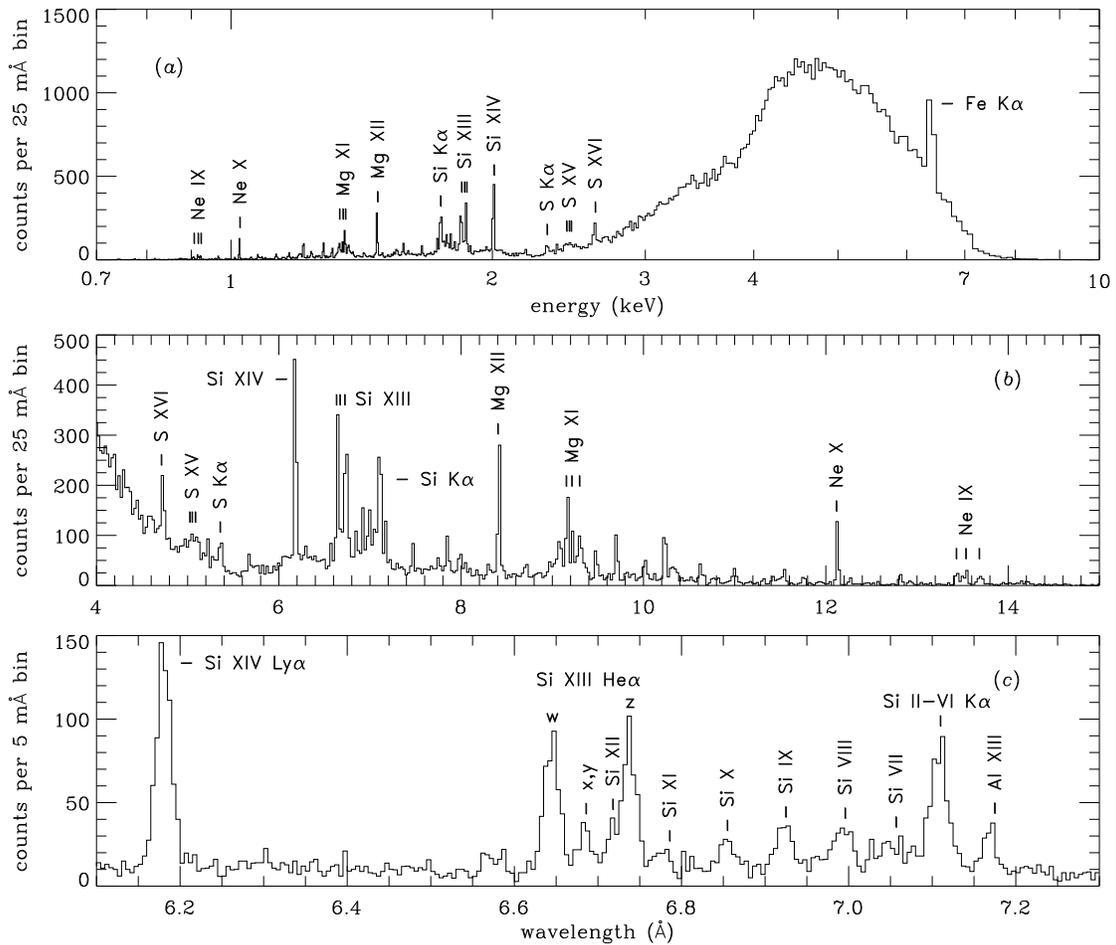} 
\caption{Three views of the {\it Chandra\/} HETG Medium Energy Grating
count spectrum of Vela X-1 in eclipse, with the strongest emission
features labeled:
({\it a\/}) the broad 0.7--10 keV bandpass,
({\it b\/}) the line-rich 4--15~\AA \ bandpass, and
({\it c\/}) the 6.1--7.3~\AA \ bandpass containing emission lines
from H- and He-like Si as well as fluorescence emission features
from L-shell Si.}
\end{figure}

\section{Vela X-1}

The prototypical detached HMXB Vela X-1 has been studied extensively
in nearly every waveband, particularly in X-rays, since its discovery
as an X-ray source during a rocket flight four decades ago. It
consists of a B0.5 Ib supergiant and a magnetic neutron star in an
8.964-day orbit. From an X-ray spectroscopic point of view, Vela X-1
distinguished itself in 1994 when Nagase et al.\ \cite{nag94}, using
{\it ASCA\/} SIS data, showed that, in addition to the well-known
6.4 keV Fe K$\alpha $ emission line, the eclipse X-ray spectrum is
dominated by recombination lines and continua of H- and He-like Ne,
Mg, Si, S, Ar, and Fe. These data were subsequently modeled in detail
by Sako et al.\ \cite{sak99}, using a kinematic model in which the
photoionized wind was characterized by the ionization parameter
$\xi\equiv L_{\rm x}/nR^2$, where $R$ is the distance from the neutron
star and $n$ is the number density, given by the mass-loss rate and
velocity law of an undisturbed CAK wind [specifically, $n=\Mdot _{\rm
w}/4\pi\mu v(r)r^2$, where $\mu $ is the mean atomic weight and
$v(r)=v_0+v_\infty (1-R_\star /r)^\beta $]. Vela X-1 was subsequently
observed with the {\it Chandra\/} High Energy Transmission Grating
(HETG) in 2000 for 30 ks in eclipse \cite{sch02} and in 2001 for 85,
30, and 30 ks in eclipse and at binary phases 0.25 and 0.5,
respectively \cite{gol04,wat06}; three views of the line-rich eclipse
spectrum are shown in Figure 1. Watanabe et al.\ \cite{wat06}, using
assumptions very similar to those of Sako et al.\ and a Monte Carlo
radiation transfer code, produced a global model of Vela X-1 that
simultaneously fits the HETG spectra from the
three binary phases with a wind mass-loss rate $\Mdot_{\rm w}\approx 
2\times 10^{-6}~\Msun~\yr $ and terminal velocity $v_\infty = 1100~\rm
km~s^{-1}$. One of the failures of this model was the velocity shifts
of the emission lines between eclipse and binary phase 0.5, which were
observed to be $\Delta v\approx 400$--$500~\rm km~s^{-1}$, while the
model simulations predicted $\Delta v\sim 1000~\rm km~s^{-1}$. In
order to resolve this discrepancy, Watanabe et al.\ performed a
1-dimensional calculation to estimate the wind velocity profile along
the line of centers between the two stars, accounting, in an
approximate way, for the reduction of the radiative driving due to
photoionization. They found that the velocity of the wind near the
neutron star is lower by a factor of 2--3 relative to that of an
undisturbed CAK wind, which was sufficient to explain the observations.
However, these results were not fed back into their global model to
determine the effect on the X-ray spectra. Equation 1 suggests that,
for a given wind mass-loss rate, the X-ray luminosity should be higher
by a factor of (2--$3)^4=16$--81, or, for a given X-ray luminosity,
that the wind mass-loss rate should be lower by a similar factor.

\section{Hydrodynamic Simulations}

To make additional progress in our understanding of the wind and
accretion flow of HMXBs in general and Vela X-1 in particular --- to
combine the best features of the hydrodynamic models of Blondin et
al.\ \cite{blo90,blo91,blo94,blo95} and the kinematic-spectral models
of Sako et al.\ \cite{sak99} and Watanabe et al.\ \cite{wat06} --- we
have undertaken a project to develop improved models of
radiatively-driven photoionized accretion flows, with the goal of
producing synthetic X-ray spectra that possess a level of detail
commensurate with the grating spectra returned by {\it Chandra\/} and
{\it XMM-Newton\/}. This project combines:
\begin{itemize}
\item XSTAR \cite{kal01} photoionization calculations, which provide
the heating $\Gamma $ and cooling $\Lambda $ rates of the plasma and
the ionization fractions of the various ions as a function of
temperature $T$ and ionization parameter $\xi $.
\item HULLAC \cite{bar88} emission models appropriate to X-ray photoionized
plasmas.
\item Improved models of the radiative driving of the photoionized wind.
\item FLASH \cite{fry00} two- and three-dimensional time-dependent adaptive-mesh
hydrodynamics calculations.
\item Monte Carlo radiation transport \cite{mau04}.
\end{itemize}

Radiative driving of the wind is accounted for via the force
multiplier formalism \cite{cak75}, wherein the radiative acceleration
$g_{\rm rad}=g_{\rm e}\, (1+M)$, where
$g_{\rm e}=\kappa_{\rm e}L_\star/4\pi r^2c$ is the radiative acceleration due
to electron (Thomson) scattering;
$\kappa_{\rm e}=n_{\rm e}\sigma_{\rm T}/\rho \approx 0.34~\rm
cm^2~g^{-1}$ is the Thomson opacity;
$n_{\rm e}$ is the electron number density;
$\rho $ is the mass density;
$M=\sum _{\rm lines}(F_\nu /F) (v_{\rm th}\nu /c) (1-e^{-\eta t})/t$
is the line force multiplier;
$v_{\rm th}=(2kT_\star/m_{\rm p})^{1/2}$ is the mean thermal velocity
of the protons;
$T_\star $, $L_\star$, and $F_\nu /F$ are respectively the effective
temperature, luminosity, and normalized flux distribution of the OB
star;
$t=\kappa_{\rm e}v_{\rm th}\rho (dv/dr)^{-1}$ is the optical depth parameter;
$dv/dr$ is the velocity gradient;
$\eta $ accounts for all the atomic physics (wavelengths, oscillator
strengths, statistical weights, and fractional level populations); and
the other symbols have their usual meanings. We calculated $M[T, \xi,
t]$ on a $[21\times 51 \times 41]$ lattice accounting for X-ray
photoionization and non-LTE population kinetics using HULLAC atomic
data for $2\times 10^6$ lines of $35{,}000$ energy levels of 166 ions
of the 13 most cosmically abundant elements. Compared to Stevens and
Kallman \cite{ste90}, our line force multipliers are typically higher
at low $t$ because of the larger number of lines in our models, and
lower at high $t$ because we do not assume that the level populations
are in LTE.

In addition to the usual hydrodynamic quantities, the FLASH
calculations account for:
\begin{itemize}
\item The gravity of the OB star and neutron star (although, following Ruffert
      \cite{ruf94}, for numerical reasons we softened the gravitational
      potential of the neutron star over a scale length $\epsilon=10^3\, R_{\rm
      x}=10^9$ cm).
\item Coriolis and centrifugal forces.
\item Radiative driving of the wind as a function of the local ionization
      parameter, temperature, and optical depth.
\item Photoionization and Compton heating of the irradiated wind.
\item Radiative cooling of the irradiated wind and the ``shadow wind'' behind
      the OB star.
\end{itemize}

\begin{figure}
\includegraphics[scale=0.875]{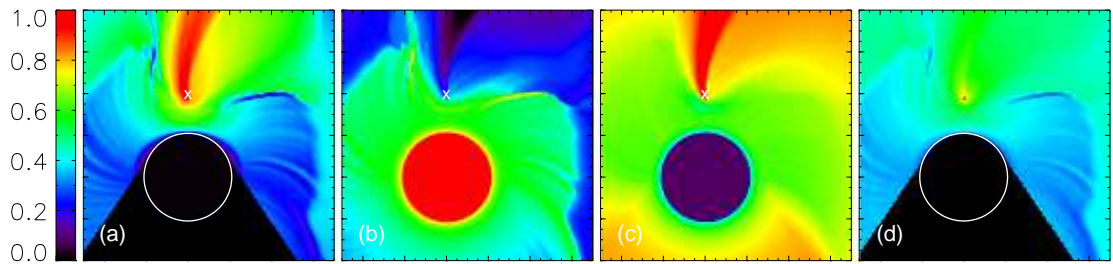} 
\caption{Color-coded maps of
({\it a\/}) $\log T\,  ({\rm K})=[4.4,8.3]$,
({\it b\/}) $\log n\,  ({\rm cm}^{-3})=[7.4,10.8]$,
({\it c\/}) $\log v\,  ({\rm km~s^{-1}})=[1.3,3.5]$, and
({\it d\/}) $\log\xi\, ({\rm erg~cm~s^{-1}})=[1.1,7.7]$ in the orbital plane
of Vela~X-1. The positions of the B star and neutron star are shown by the
circle and the ``$\times $,'' respectively. The horizontal axis $x=[-5,7]
\times 10^{12}$ cm and the vertical axis $y=[-4,8]\times 10^{12}$ cm.}
\end{figure}

\section{2D Hydrodynamic Simulations}

To demonstrate typical results of our simulations, we show in Fig.~2
color-coded maps of the log of the
({\it a\/}) temperature $T$,
({\it b\/}) density $n$,
({\it c\/}) velocity $v$, and
({\it d\/}) ionization parameter $\xi$ of a FLASH 2-dimensional
simulation in the binary orbital plane of an HMXB with parameters
appropriate to Vela X-1 (specifically, those of Sako et al.\
\cite{sak99}). In this simulation, the computational volume
$x=[-5,7]\times 10^{12}$ cm and $y=[-4,8]\times 10^{12}$ cm, the B
star is centered at $[x,y]=[0,0]$, the neutron star is centered at
$[x,y]=[0,3.7\times 10^{12}]$ cm, and $128\times 128=16{,}384$
computational cells were used for a spatial resolution of $\Delta
l=9.4\times 10^{10}$ cm. At the time step shown ($t = 100$ ks), the
relatively slow ($v \approx 400~\rm km~s^{-1}$)\footnote{Note that
this velocity reproduces the value that Watanabe et al.\ found was
needed to match the velocity of the emission lines in the {\it
Chandra\/} HETG spectra of Vela X-1.} irradiated wind has reached just
$\sim 2$ stellar radii from the stellar surface. The various panels
show (1) the effect of the Coriolis and centrifugal forces, which
cause the flow to curve clockwise, (2) the cool, fast wind behind the
B star, (3) the hot, slow, irradiated wind, (4) the hot, low-density,
high-velocity flow downstream of the neutron star, and (5) the bow
shock and two flanking shocks formed where the irradiated wind
collides with the hot disturbed flow in front and downstream of the
neutron star.

\begin{figure}
\includegraphics[scale=0.875]{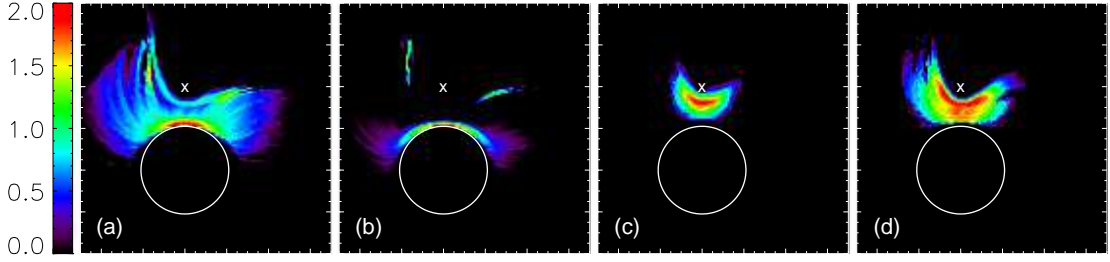} 
\caption{Similar to Fig.~2, but for the log of the X-ray emissivity of
({\it a\/}) \ion{Si}{XIV}  Ly$\alpha $,
({\it b\/}) \ion{Si}{XIII} He$\alpha $,
({\it c\/}) \ion{Fe}{XXVI} Ly$\alpha $, and
({\it d\/}) \ion{Fe}{XXV}  He$\alpha $ in the orbital plane of Vela~X-1.
In each case, two orders of magnitude are plotted.}
\end{figure}

Given these maps, it is straightforward to determine where in the binary the
X-ray emission originates. To demonstrate this, we show in Fig.~3
color-coded maps of the log of the emissivity of
({\it a\/}) \ion{Si}{XIV}  Ly$\alpha $,
({\it b\/}) \ion{Si}{XIII} He$\alpha $,
({\it c\/}) \ion{Fe}{XXVI} Ly$\alpha $, and
({\it d\/}) \ion{Fe}{XXV}  He$\alpha $.
The gross properties of these maps are consistent with Fig.~24 of
Watanabe et al., but they are now (1) quantitative rather than
qualitative and (2) specific to individual transitions of individual
ions. The maps also capture features that otherwise would not have
been supposed, such as the excess emission in the H- and He-like Si
lines downstream of the flanking shocks. Combining these maps with the
velocity map (Fig.~2{\it c\/}), these models make very specific
predictions about the intensity of the emission features, where they
originate, and their velocity widths and amplitudes as a function of
binary phase.

\section{3D Hydrodynamic Simulations}

Having developed and extensively tested our software on various
2-dimensional simulations, we next calculated a more limited number
of 3-dimensional simulations of an HMXB with parameters appropriate
to Vela X-1. Figure 4 shows,
for one of these simulations, the log of the number density on three
orthogonal planes passing through the computational volume, which
spans $\Delta x = 12\times 10^{12}$ cm, $\Delta y = 16\times 10^{12}$
cm, and $\Delta z = 8\times 10^{12}$ cm using $192\times 256\times
128=6.3\times 10^6$ computational cells, for a spatial resolution
$\Delta l = 6.3\times 10^{10}$ cm. In order to compare our hydrodynamic
calculations to the kinematic models of Sako et al.\ \cite{sak99}
and Watanabe et al.\ \cite{wat06}, we also calculated, on the same
3-dimensional grid, using the parameters of Watanabe et al., the
velocity $v(r)=v_0+v_\infty (1-R_\star /r)^\beta $, density
$n= \Mdot _{\rm w}/4\pi\mu v(r)r^2$, and ionization parameter
$\xi\equiv L_{\rm x}/nR^2$ for an assumed undisturbed CAK wind,
assuming that the temperature $T(\xi )$ is that of a photoionized
plasma in thermal equilibrium: $\Gamma (T,\xi)=\Lambda (T,\xi)$.
Figure 1 of Akiyama et al.\ \cite{aki07} shows color-coded maps in
the binary orbital plane of the density and temperature and the
\ion{Si}{XIV} and \ion{Fe}{XXVI} ion abundances for the CAK and
FLASH models, emphasizing the dramatic differences between them.

\begin{figure}
\includegraphics[height=2.0in]{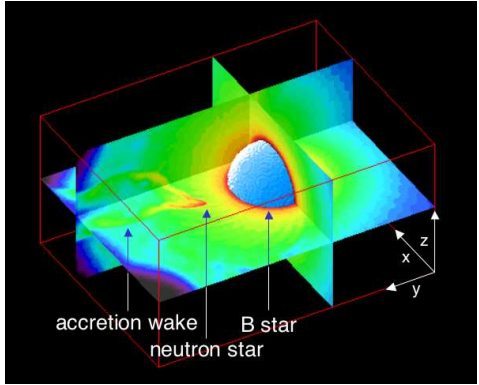} 
\caption{Color-coded maps of $\log n\, ({\rm cm}^{-3})=[7.4,12.1]$
on three orthogonal planes from a FLASH 3D simulation of Vela X-1,
showing the B star ({\it blue sphere\/}), neutron star, and accretion
wake. The computational volume $x=[-6,6]\times 10^{12}$ cm, $y=[-6,10]
\times 10^{12}$ cm, and $z=[-4,4]\times 10^{12}$ cm.}
\end{figure}

\section{Synthetic X-ray Spectra}

In the next step in this process, we fed the physical parameters of
the CAK and FLASH models into our Monte Carlo radiation transfer code
and followed the spatial and spectral evolution of photons launched
from the surface of the neutron star until they are either destroyed
or escape the computational volume. As detailed by Mauche et al.\
\cite{mau04}, the Monte Carlo code accounts for Compton and
photoelectric opacity of 446 subshells of 140 ions of the 12 most
abundant elements. Following photoabsorption by K-shell ions, we
generate radiative recombination continua (RRC) and recombination line
cascades in a probabilistic manner using the recombination cascade
calculations described by Sako et al.\ \cite{sak99}, with the shapes
of the RRCs determined by the functional form of the photoionization
cross sections and the local electron temperature. Fluorescence 
emission from L-shell ions is ignored, since it is assumed to be
suppressed by resonant Auger destruction (however, see Fig.\ 
1({\it c}) and Liedahl \cite{lie05}).

The synthetic spectra resulting from the Monte Carlo simulations are
shown in Fig.~5. The upper histogram is the assumed (featureless)
power-law spectrum emitted by the neutron star, while the synthetic
spectra for the CAK and FLASH physical models are shown by the middle
and lower histograms, respectively.
While it is too early in the development of our program to draw any
firm conclusions from these spectra (or from the differences between
them), they demonstrate the type of results produced by our Monte
Carlo code --- X-ray spectra rich in emission lines and RRCs --- and
the type of differences between simple CAK and detailed hydrodynamic
models, highlighting the importance of the details of the
3-dimensional model to the emitted spectra.

\begin{figure}
\includegraphics[]{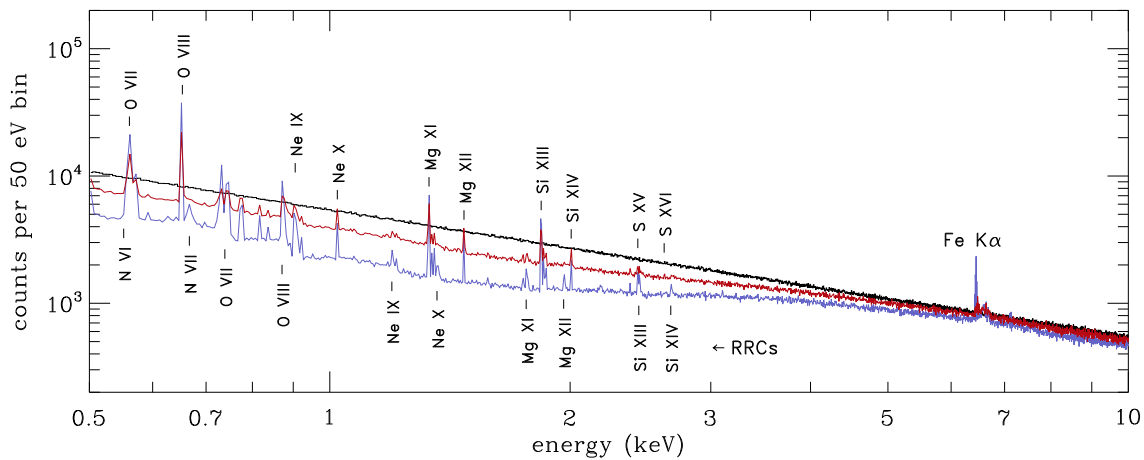} 
\caption{Monte Carlo X-ray spectra of the CAK ({\it red\/}) and
FLASH ({\it blue\/}) models of Vela X-1 for the assumed power law
neutron star X-ray spectrum ({\it black\/}). The H- and He-like
ions responsible for the strongest emission lines and radiative
recombination continua are indicated above and below the spectra,
respectively.}
\end{figure}

We close by noting that Fig.~5 plots {\it global\/} X-ray spectra,
irrespective of viewing angle or occultation by the B star, although
it is straightforward to window the output of the Monte Carlo code
to produce synthetic spectra for a specific viewing direction (for
a given binary inclination and binary phase) or for a given binary
inclination and a range of binary phases, and hence allow direct
comparisons to actual data. The next step in our program is a careful
comparison of such synthetic spectra to the grating spectra of Vela
X-1 and other HMXBs returned by {\it Chandra\/} and {\it
XMM-Newton\/}. In this way, it will be possible to better constrain
the various parameters of the winds of HMXBs, such as the mass-loss
rate $\Mdot _{\rm w}$, terminal velocity $v_\infty $, velocity profile
$v(r)$, and elemental abundances of the OB star --- parameters that
bear on such fundamental questions as the long-term evolution
of these binaries and the chemical enrichment of the interstellar medium.


\begin{theacknowledgments}

This work was performed under the auspices of the U.S.\ Department
of Energy by Lawrence Livermore National Laboratory under Contract
DE-AC52-07NA27344. Support for this work was provided by the
Laboratory Directed Research and Development Program at LLNL under
project tracking code 05-ERD-044 and by the National Aeronautics and
Space Administration under Agreement No.\ NN07AF91I issued through the
Astrophysics Theory Program. T.\ Plewa's contribution to this work was
supported in part by the U.S.\ Department of Energy under Grant No.\
B523820 to the Center for Astrophysical Thermonuclear Flashes at
the University of Chicago. The FLASH software used in this work was
developed in part by the DOE-supported ASC/Alliance Center for
Astrophysical Thermonuclear Flashes at the University of Chicago.

\end{theacknowledgments}




\begin{thebibliography}{99}

\bibitem{cak75}
         J.~I.~Castor, D.~C.~Abbott, and R.~I.~Klein,
         \emph{Ap.J.}, \textbf{195}, 157 (1975), CAK.
\bibitem{hoy39}
         F.~Hoyle and R.~A.~Lyttleton,
         \emph{Proc.~Cam.~Phil.~Soc.}, \textbf{35}, 405 (1939).
\bibitem{bon52} Bondi, H.,
         \emph{M.~N.~R.~A.~S.}, \textbf{112}, 195 (1952)
\bibitem{sha76}
         S.~L.~Shapiro and A.~P.~Lightman,
         \emph{Ap.J.}, \textbf{204}, 555 (1976).
\bibitem{hat77}
         S.~Hatchett and R.~McCray,
         \emph{Ap.J.}, \textbf{211}, 552 (1977).
\bibitem{mcc84}
         R.~McCray, T.~R.~Kallman, J.~I.~Castor, and G.~L.~Olson,
         \emph{Ap.J.}, \textbf{282}, 245 (1984).
\bibitem{ste90}
         I.~R.~Stevens and T.~R.~Kallman,
         \emph{Ap.J.}, \textbf{365}, 321 (1990).
\bibitem{blo90}
         J.~M.~Blondin, T.~R.~Kallman, B.~A.~Fryxell, and R.~E.~Taam,
         \emph{Ap.J.}, \textbf{356}, 591 (1990).
\bibitem{blo91}
         J.~M.~Blondin, I.~R.~Stevens, and T.~R.~Kallman,
         \emph{Ap.J.}, \textbf{371}, 684 (1991).
\bibitem{blo94}
         J.~M.~Blondin,
         \emph{Ap.J.}, \textbf{435}, 756 (1994).
\bibitem{blo95}
         J.~M.~Blondin and J.~W.~Woo,
         \emph{Ap.J.}, \textbf{445}, 889 (1995).
\bibitem{nag94}
         F.~Nagase, G.~Zylstra, T.~Sonobe, T.~Kotani, H.~Inoue, and J.~Woo,
         \emph{Ap.J.}, \textbf{436}, L1 (1994).
\bibitem{sak99}
         M.~Sako, D.~A.~Liedahl, S.~M.~Kahn, and F.~Paerels,
         \emph{Ap.J.}, \textbf{525}, 921 (1999).
\bibitem{sch02}
         N.~S.~Schulz, C.~R.~Canizares, J.~C.~Lee, and M.~Sako,
         \emph{Ap.J.}, \textbf{564}, L21 (2002).
\bibitem{gol04}
         G.~Goldstein, D.~P.~Huenemoerder, and D.~Blank,
         \emph{A.J.}, \textbf{127}, 2310 (2004).
\bibitem{wat06}
         S.~Watanabe, et al.,
         \emph{Ap.J.}, \textbf{651}, 421 (2006).
\bibitem{kal01}
         T.~Kallman and M.~Bautista,
         \emph{Ap.J.S.}, \textbf{133}, 221 (2001).
\bibitem{bar88}
         A.~Bar-Shalom, M.~Klapisch, and J.~Oreg,
         \emph{Phys.~Rev.}, \textbf{A38}, 1773 (1988).
\bibitem{fry00}
         B.~Fryxell, et al.,
         \emph{Ap.J.S.}, \textbf{131}, 273 (2000).
\bibitem{mau04}
         C.~W.~Mauche, D.~A.~Liedahl, B.~F.~Mathiesen, M.~A.~Jimenez-Garate,
         and J.~C.~Raymond,
         \emph{Ap.J.}, \textbf{606}, 168 (2004).
\bibitem{ruf94}
         M.~Ruffert,
         \emph{Ap.J.}, \textbf{427}, 342 (1994).
\bibitem{aki07}
         S.~Akiyama, C.~W.~Mauche, D.~A.~Liedahl and T.~Plewa,
         \emph{Bull.~A.~A.~S.}, \textbf{39}, \#20.03 (2007)
\bibitem{lie05}
         D.~A.~Liedahl,
         ``Resonant Auger Destruction and Iron K$\alpha $  Spectra in
         Compact X-ray Sources,
         in \emph{X-ray Diagnostics of Astrophysical Plasmas: Theory,
         Experiment, and Observation}, edited by R.~Smith,
         AIP Conference Proceedings 774, 2005, p. 99.

\end{thebibliography}
\end{document}